\documentclass[conference]{IEEEtran}
\usepackage{fancyhdr}
\IEEEoverridecommandlockouts
\usepackage{cite}
\usepackage{amsmath,amssymb,amsfonts}
\usepackage{algorithmic}
\usepackage{graphicx}
\usepackage{textcomp}
\usepackage{relsize}
\usepackage{xcolor}
\def\BibTeX{{\rm B\kern-.05em{\sc i\kern-.025em b}\kern-.08em
    T\kern-.1667em\lower.7ex\hbox{E}\kern-.125emX}}

\newif\ifmicrotype
\newif\ifarxiv
\arxivtrue
\microtypetrue

\ifmicrotype
\usepackage[%
activate={true,nocompatibility},%
final,%
tracking=true,%
kerning=true,%
spacing=true,%
stretch=10,%
shrink=30]{microtype}
\microtypecontext{spacing=nonfrench}
\SetTracking{encoding={*}, shape=sc}{0}
\fi

\usepackage{hyperref}
\usepackage{soul}

\makeatletter
\newcommand{\MPI}[1]{\mbox{\small\texttt{MPI\_\@capitalizefirst#1\@nil}}}
\newcommand{\alltoall}[0]{all-to-all\xspace}
\newcommand{\Alltoall}[0]{All-to-all\xspace}
\newcommand{\ErdosRenyi}[0]{Erd\H{o}s-R\'{e}nyi}

\newcommand{\Cpp}{C\nolinebreak[4]\hspace{-.05em}\raisebox{.13ex}{\smaller{\smaller \bf ++}}\xspace}

\newcommand{\frage}[1]{#1}
\renewcommand{\frage}[1]{}

\renewcommand{\nu}[1]{\textcolor{blue}{\frage{[nu:#1]}}}

\def\@capitalizefirst#1#2\@nil{\uppercase{#1}#2}
\makeatother

\usepackage{xspace}
\makeatletter
\newcommand{\kamping}{%
  \mbox{KaMPIng}\xspace%
}
\makeatother

\definecolor{my-dark-red}{RGB}{183, 28, 28}
\definecolor{my-red}{RGB}{244,67,54}
\definecolor{my-pink}{RGB}{233,30,99}
\definecolor{my-purple}{RGB}{156,39,176}
\definecolor{my-deep-purple}{RGB}{103,58,183}
\definecolor{my-indigo}{RGB}{63,81,181}
\definecolor{my-blue}{RGB}{33,150,243}
\definecolor{my-light-blue}{RGB}{3,169,244}
\definecolor{my-cyan}{RGB}{0,188,212}
\definecolor{my-teal}{RGB}{0,150,136}
\definecolor{my-green}{RGB}{76,175,80}
\definecolor{my-light-green}{RGB}{139,195,74}
\definecolor{my-lime}{RGB}{205,220,57}
\definecolor{my-yellow}{RGB}{255,235,59}
\definecolor{my-amber}{RGB}{255,193,7}
\definecolor{my-orange}{RGB}{255,152,0}
\definecolor{my-deep-orange}{RGB}{255,87,34}
\definecolor{my-brown}{RGB}{121,85,72}
\definecolor{my-grey}{RGB}{158,158,158}
\definecolor{my-blue-grey}{RGB}{96,125,139}
\definecolor{my-lipics-grey}{rgb}{0.6,0.6,0.61}

\usepackage[per-mode=symbol,binary-units=true]{siunitx}[=v2]
\usepackage{listings}
\usepackage{pgfplots}
\pgfplotsset{compat=1.16}

\usepackage{pgf}

\everymath=\expandafter{\the\everymath\displaystyle}

\usepackage{tablefootnote}
\usepackage{threeparttable}
\usepackage{pifont}

\usepackage[capitalize]{cleveref}
\crefname{figure}{\figurename}{\figurename}
\Crefname{figure}{\figurename}{\figurename}

\usepackage{placeins}
\usepackage{tabularx}
\usepackage{booktabs}
\usepackage{url}
\usepackage{csquotes}
\usepackage{wrapfig}

\begin{document}
\ifarxiv
  \nocite{Uhl2024}
\fi
\bstctlcite{limitauthorsto4}

\title{KaMPIng: Flexible and (Near) Zero-Overhead \Cpp Bindings for MPI
  \thanks{
        \begin{wrapfigure}{R}{.33\columnwidth}
      \vspace{-1.25\baselineskip}
      \includegraphics[width=.33\columnwidth]{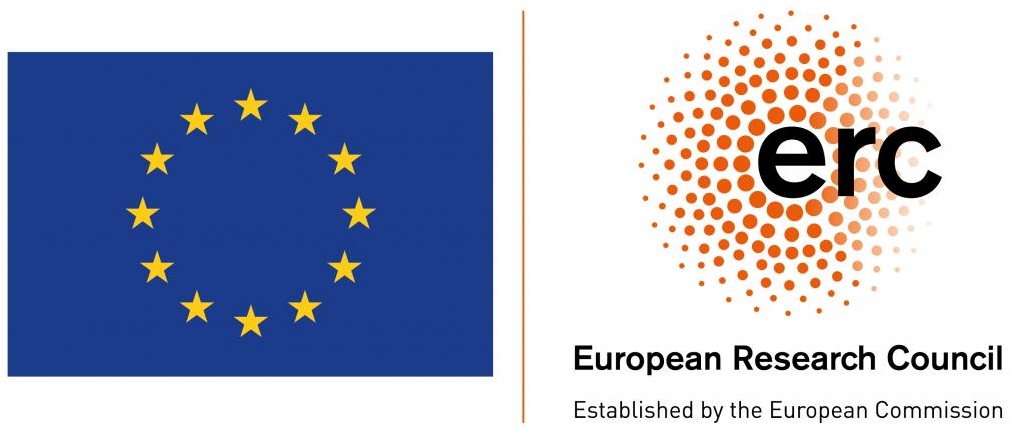}
    \end{wrapfigure}	

    This project has received funding from the European Research
    Council (ERC) under the European Union’s Horizon 2020 research and
    innovation programme (grant agreement No. 882500).

  }
  \thanks{
    This work was supported by a grant from the Ministry of Science, Research and the Arts of Baden-Württemberg (Az: 33-7533.-9-10/20/2) to Peter Sanders and Alexandros Stamatakis.
  }
}

\author{
  \IEEEauthorblockN{
    Tim Niklas Uhl\IEEEauthorrefmark{1},
    Matthias Schimek\IEEEauthorrefmark{1},
    Lukas Hübner\IEEEauthorrefmark{1}\IEEEauthorrefmark{2},
    Demian Hespe\IEEEauthorrefmark{3}\\
    Florian Kurpicz\IEEEauthorrefmark{1},
    Daniel Seemaier\IEEEauthorrefmark{1},
    Christoph Stelz\IEEEauthorrefmark{1}
    and Peter Sanders\IEEEauthorrefmark{1}
  }

  \IEEEauthorblockA{
    \IEEEauthorrefmark{1}
    \textit{Institute of Theoretical Informatics} \\
    \textit{Karlsruhe Institute of Technology}\\
    Karlsruhe, Germany\\
    \{uhl, schimek, huebner, kurpicz, daniel.seemaier, sanders\}@kit.edu,
  }
  \IEEEauthorblockA{
    \IEEEauthorrefmark{2}
    \textit{Heidelberg Institute for Theoretical Studies} \\
    Heidelberg, Germany
  }
  \IEEEauthorblockA{
    \IEEEauthorrefmark{3}
    \textit{Independent}\\
    demian.hespe@outlook.com
  }
}

\maketitle
\thispagestyle{fancy}
\lhead{}
\rhead{}
\chead{}
\lfoot{\footnotesize{
    \ifarxiv
      \vspace{-3em}
      Published at
    \fi
    SC24, November 17-22, 2024, Atlanta, Georgia, USA%
    \ifarxiv
      , publisher's version avaible under DOI \href{https://doi.ieeecomputersociety.org/10.1109/SC41406.2024.00050}{10.1109/SC41406.2024.00050} \cite{Uhl2024}
      \newline
      \copyright~2024 IEEE.
      Personal use of this material is permitted.
      Permission from IEEE must be obtained for all other uses, in any current or future media, including reprinting/republishing this material for advertising or promotional purposes, creating new collective works, for resale or redistribution to servers or lists, or reuse of any copyrighted component of this work in other works.
    \else
      \newline 979-8-3503-5291-7/24/\$31.00 \copyright 2024 IEEE
    \fi
  }}
\rfoot{}
\cfoot{}
\renewcommand{\headrulewidth}{0pt}
\renewcommand{\footrulewidth}{0pt}

\begin{abstract}
	The Message-Passing Interface (MPI) and \Cpp form the backbone of high-performance computing, but MPI only provides C and Fortran bindings.
	While this offers great language interoperability, high-level programming languages like \Cpp make software development quicker and less error-prone.

	We propose novel \Cpp language bindings that cover all abstraction levels from low-level MPI calls to convenient STL-style bindings, where most parameters are inferred from a small subset of parameters, by bringing named parameters to \Cpp.
	This enables rapid prototyping and fine-tuning runtime behavior and memory management.
	A flexible type system and additional safety guarantees help to prevent programming errors.
	
        By exploiting \Cpp's template metaprogramming capabilities, this has (near) zero overhead, as only required code paths are generated at compile time.

	We demonstrate that our library is a strong foundation for a future distributed standard library using multiple application benchmarks, ranging from text-book sorting algorithms to phylogenetic interference.
\end{abstract}


\lstset{language=c++, basicstyle=\ttfamily\small}

\section{Introduction}
The first version of the Message-Passing Interface (MPI) was proposed by the Message-Passing Interface Forum in 1994~\cite{Forum2023} with the goal of standardizing a portable, flexible, and efficient standard for message-passing.
Today, it is the backbone of most HPC applications.
While the majority of them are written in \Cpp~\cite{Laguna2019}, MPI's syntax and semantics are designed around C and Fortran.
While this allows for calling MPI from \Cpp code, the semantics do not fit well with modern \Cpp language features.
This makes developing MPI applications in \Cpp unintuitive and error-prone~\cite{Ruefenacht2021}.

MPI 2.0 (1997) introduced \Cpp bindings, which were deprecated with MPI 2.2 (2009).
With version 3.0 (2012), the bindings have been removed entirely, because they only added minimal functionality over the C bindings while adding significant maintenance complexity to the MPI specification~\cite{Forum2009}.

Since then, there have been various efforts in designing new \Cpp interfaces.
Notable libraries include \emph{Boost.MPI}~\cite{Gregor2007}, the MPI bindings by Demiralp et~al.~\cite{Demiralp2023}, and \emph{MPL}~\cite{Bauke2015}, which has recently been considered as a starting point for new \Cpp language bindings by the newly formed MPI working group on language bindings~\cite{Ghosh2021}.

Previous bindings focussed on making MPI's C interface compatible with \Cpp features such as templates, STL containers and object orientation, and providing some sensible defaults, but this resulted in each library choosing its own level of abstraction:
They either provide a high-level interface, while often sacrificing performance~\cite{Ghosh2021}, or stay close to MPI's C interface, still requiring large amounts of boilerplate code.
Additionally, these previous libraries do not work well with move semantics and common practices proposed by the \Cpp core guidelines, e.g., returning results by value instead of using C-style in/out parameters~\cite[F.20]{CppCoreGuidelines}.
Despite the previous efforts, designing a better \Cpp MPI interface is an open problem actively discussed in the MPI forum~\cite{Forum2020}.

Therefore, we propose the C++ MPI binding library \textbf{\kamping} (Karlsruhe MPI next generation)\footnote{\url{https://github.com/kamping-site/kamping}}, which we recently announced briefly at SPAA~\cite{Hespe2024}.
Its main goal is to cover the complete range of abstraction levels over MPI calls as shown in \cref{fig:kamping-intro}:
\kamping enables rapid prototyping, relying on its sensible defaults, as well as engineering highly-tuned distributed code by making use of the library's flexible interface.
By using C++ template metaprogramming techniques this comes without introducing significant overhead compared to MPI's plain C bindings.

Parameters of MPI calls can either be provided directly by the user or are computed by \kamping.
It further offers complete control over memory allocations.

It helps to reduce common sources of programming errors by employing compile-time error checking and prevents invalid memory access for non-blocking communication by introducing an ownership model.
A flexible type system supports type safety by generating type definitions at compile time and enables serialization when needed.
Because all this is achieved using template metaprogramming, only the code paths programmers would have to write themselves are instantiated, while the rest can be eliminated at compile-time, which makes these new bindings near zero-overhead.

Currently, \kamping supports the most commonly used MPI features~\cite{Laguna2019}, namely (non-blocking) point-to-point communication and collective communication.
Moreover, its core architecture is designed with the rest of the MPI standard in mind, facilitating a straightforward implementation of it in the future.

We highlight the flexibility of our library by discussing a variety of different application benchmarks, including both simple examples and more complex application scenarios.
This includes applications in sorting, text processing, graph search and partitioning, and phylogenetic interference.
For example, \kamping allows us to implement the Prefix Doubling algorithm for suffix array construction~\cite{Manber1993} using less than 200 lines of code and to entirely remove the custom MPI abstraction layer of RaxML-NG~\cite{raxml-ng}.

\begin{figure}[t]
	\begin{lstlisting}[
basicstyle=\ttfamily\scriptsize,
emph={[2]send_buf, send_counts, recv_counts_out, recv_displs_out},
emphstyle={[2]\color{gray}},
commentstyle=\textcolor{my-dark-red},
emph={[3]vector, std, exclusive_scan, move},
emphstyle={[3]\color{my-purple}},
escapechar=!
]
std::vector<double> v = {...};
// KaMPIng allows concise code
// with sensible defaults ... (1)
auto v_global = comm.allgatherv(send_buf(v));

// ... or detailed tuning of each parameter (2)
std::vector<int> rc;
auto [v_global, rcounts, rdispls] = comm.allgatherv(
 send_buf(v), //(3)
 recv_counts_out<resize_to_fit/*(6)*/>(std::move(rc)), //(4)
 recv_displs_out() //(5)
);
  \end{lstlisting}
  \caption{\kamping offers a high-level easy-to-use interface (1) and full control over each parameter (2). Data types and buffer sizes can be automatically inferred (3). Arguments allow passing by reference or by value and transferring data ownership via move semantics (4). Out-parameters allow controlling which default-computed parameters are returned to the user (5). Resize policies allow controlling memory allocation (6).}
  \label{fig:kamping-intro}
\end{figure}
\subsection*{Our Contributions}
In the following, we briefly list our contributions:
\begin{itemize}
\item A new approach to parameter handling, allowing for a flexible computation of default parameters and fine-grained control over memory management for highly engineered distributed applications.
\item Safety for non-blocking MPI communication calls by design.
\item Reduction of verbosity and error-proneness of MPI code.
\item (Near) zero overhead compared to using plain MPI.
\item Compile-time error checking with human-readable error messages.
\item Demonstration of real-world applicability using a large variety of application benchmarks.
\item First steps towards integrated and extensible general algorithmic building blocks for distributed computing such as specialized collectives for sparse and low-latency irregular personalized communication.
\end{itemize}

The remainder of the paper is organized as follows.
We begin by providing an overview of existing MPI (\Cpp) language bindings in \cref{sec:related-work}.
In \cref{sec:overview-and-design}, we present \kamping's core features like parameter handling, type deduction and serialization, handling of non-blocking communication, and our approach to ensure easy expandability.
We then evaluate the applicability of \kamping in multiple different real-world application benchmarks in \cref{sec:evaluation}.
Finally, we present first examples for general distributed computing algorithmic building blocks in \cref{sec:general-building-blocks} and conclude by discussing possibilities for future work in \cref{sec:conclusion}.

\section{Related Work}
\label{sec:related-work}
Since the removal of \Cpp bindings from the MPI standard, there has been continuous third-party effort on designing \Cpp language bindings for MPI.
The topic is also actively discussed in the MPI forum~\cite{Forum2020},
where the following desired features were prominently mentioned:
\begin{itemize}
\item mapping of variables to types, type safety
\item safety guarantees for non-blocking communication
\item returning data by value
\item support for unstructured send data, i.e.\ a mapping of communication partners to data buffers
\item a strong debug mode
\item reduction via lambda
\item good interaction with C++, especially move semantics and ranges
\end{itemize}

In the following, we discuss notable libraries and summarize their design and feature set.

\paragraph*{Boost.MPI~\cite{Gregor2007}}
This was the first library to enable automatic data type inference and facilitate integration with the STL by also supporting \lstinline{std::vector} for input and output, and not only raw pointers.
Vectors are automatically resized to fit the received data, which prevents invalid memory accesses, but leads to hidden allocation.
It supports custom data types by constructing appropriate data types if possible or resorting to serialization.
To enable this, users must provide a serialization function compatible with \emph{Boost.Serialization}, specifying all members explicitly. This requires the user to ensure that data type definitions and their serializations remain synchronized.
The data types are managed by a global type pool which is queried for each communication call, which ensures proper resource cleanup.
The adoption of Boost.MPI has been hindered by its tight coupling with other Boost libraries and the fact that it performs implicit serialization if data types are not directly supported by MPI~\cite{Ghosh2021}.
Additionally, it is the only MPI library considered here that is not header-only.
While this is usually not a problem with modern build tools, it becomes especially tedious when switching between MPI implementation, as it requires a separate build of Boost.MPI for each, because the MPI standard does not enforce ABI compatibility.
Besides \kamping, it is the only library that supports mapping STL functors such as \lstinline{std::plus} to the corresponding built-in MPI constant for reduction operations (which may enable optimization by the MPI implementation) and writing custom reduction operations using a simple lambda.
If an MPI error occurs, an exception is thrown.
Boost.MPI does not provide any bindings for \MPI{alltoallv}.
It has not been actively maintained since 2008 and therefore only supports MPI-1.1 features.
There also exists \emph{boost-mpi3}~\cite{Correa2018}, a rewrite of Boost.MPI developed independently of the Boost project but following its design principles.
It aims to extend Boost.MPI to the feature set of MPI-3 and \Cpp-17.
While incorporating support for iterators, MPI one-sided communication and MPI shared memory, it does not substantially improve upon Boost.MPI's design, but closely mirrors MPI's C interface.

\paragraph*{MPL~\cite{Bauke2015}}
Since 2015 Bauke maintains \emph{MPL}, a library providing MPI bindings targeting \Cpp-17.

MPL introduces a powerful type system based on so-called \emph{layouts} to programmatically construct views over chunks of contiguous memory, which can be converted to MPI data types.
While this allows for flexible communication in scientific computations, such as halo exchanges, it requires a lot of code for irregular communication patterns often found in discrete algorithms.
MPL does not expose the underlying native MPI representation of constructs such as communicators, and types directly, which complicates the iterative adaption of existing MPI code to MPL.
MPL offers support for custom data types, which requires defining a matching layout but has no serialization support.
It currently does not support error handling.

Recently, MPL has been considered by the members of the newly formed MPI working group on language bindings as a starting point for new \Cpp bindings~\cite{Ghosh2021}.
While the authors highlight the simplified interface, which offers abstraction with default parameters using overloads, they also show that MPL incurs overheads for variable-size collectives.
This is because it does not use the corresponding collective operations directly by passing appropriate counts/displacements, but constructs custom derived data types with displacements.
Therefore, operations like \lstinline{gatherv} call \MPI{alltoallw} internally, which is costly in some MPI implementations and limits scalability~\cite{Ghosh2021}.

\paragraph*{RWTH-MPI~\cite{Demiralp2023}}
Demiralp~et~al. recently introduced another modern \Cpp interface, which we call RWTH-MPI in the following.
They offer full STL support for send and receive buffers.
For each communication call, they provide various overloads using different abstraction levels, which often allow the omission of send or receive counts, and RWTH-MPI employs additional communication to compute them.
Automatic resizing of the receive buffer is supported in some cases and can be disabled.
For custom types that support reflection based on the PFR library~\cite{Polukhin2016}, RWTH-MPI constructs appropriate MPI data types automatically.
A customization point for mapping \Cpp types to type definitions for MPI is available but does not provide automatic type management.
Using types with dynamic run time sizes is not supported.

While RWTH-MPI covers the complete MPI standard, large parts directly mirror the C interface without providing additional convenience or safety guarantees.

\paragraph*{Beyond \Cpp}
There also exist MPI bindings for other high-level programming languages.
The Python bindings provided by \emph{mpi4py}~\cite{Dalcin2008,Dalcin2021} may be considered the most mature third-party bindings, as they have been actively developed for over a decade.
It enables the transparent sending of Python objects using serialization.
While this introduces additional overheads, using mpi4py in conjunction with NumPy arrays allows performance on par with a native C implementation~\cite{Dalcin2008}.
They provide default values for certain parameters, but only when no additional communication is necessary for that.
Finally, rsmpi~\cite{Steinbusch2015} introduces idiomatic Rust bindings for MPI.
While they make using MPI from Rust more ergonomic than using the C interface, the missing default parameters require the writing of much boilerplate code.

\section{\kamping: Overview and Design}
\label{sec:overview-and-design}
Realizing a \Cpp interface that is both easy to use and highly flexible, requires building a library from scratch, improving upon the design of its precursors.
Similar to most previous bindings, \kamping represents MPI objects such as communicators, requests and statuses as classes and operations on them as member functions.
Automatic creation/destruction of MPI objects is achieved using \emph{RAII}~\cite[E.6]{CppCoreGuidelines}\cite{Stroustrup1994} (Resource Acquisition Is Initialization) which is a commonly used \Cpp idiom.
Also, \Cpp data types are mapped to MPI types at compile time, which prevents type-matching errors.
STL containers that allow access to the underlying contiguous memory are directly supported, i.e., every container that models the \lstinline{std::contiguous_range} concept.
Raw pointers are supported via \lstinline{std::span} as proposed in the \Cpp core guidelines \cite[I.13]{CppCoreGuidelines}.
One of \kamping's distinct features is parameter handling.
A common source of programming errors in MPI stems from the complexity of the interface of many communication calls.
In particular, variable collective operations (suffixed with \texttt{v}) where the amount of data transferred between each processor pair varies, require a large number of parameters.
The data to be sent or received is described in terms of a pointer to a memory region, the data type, the number of elements and its displacements.
This makes MPI calls verbose and requires programmers to often consult the documentation for required parameters and ordering.
While all parameters are necessary for full flexibility, there exist many use cases where only a small subset of explicitly provided parameters suffices and the remaining ones can be inferred from them.
How we achieve this with near zero overhead is discussed in \cref{sec:defaults}.
In \cref{sec:in-out-parameters}, \kamping's overall parameter handling concept is explained in more detail.
If the number of elements to receive is already known, it may be desirable to resize containers appropriately, but for highly-tuned applications such hidden allocation may be unfavorable.
We therefore propose a flexible allocation control in \cref{sec:memory}, which is configurable at compile time.

The missing type introspection features of C require MPI users to explicitly specify the layout of data types.
If the type declaration goes out of sync with the actual data layout, this may lead to hard-to-find errors.

In \cref{sec:types} we introduce \kamping's flexible type system which provides type-safety through compile data type construction, and offers additional versatility through support for runtime-sized types and serialization.

Non-blocking communication in MPI introduces additional sources of errors, as the user has to manually wait for the completion of operations and has to take care of not performing invalid memory accesses before an operation has finished.
To address this, we propose memory-safe abstractions that prevent illegal memory accesses through the use of \Cpp's ownership model and move semantics in \cref{sec:nonblocking}.

As the MPI standard continuously grows, \Cpp bindings also need to evolve while maintaining compatibility with existing code and MPI features not covered yet by such bindings.
A key aspect here is to keep \kamping's core small, but allow easy integration of additional features via plugins.
We discuss this in \cref{sec:expandability}.

In \cref{sec:more-safety} we describe how \kamping prevents common programming errors by providing many compile- and runtime assertions and configurable error handling, and by introducing a simplified syntax for in-place operations.
Finally, \cref{sec:implementation-details} details how \kamping is implemented internally.

\subsection{Computation of Default Parameters}
\label{sec:defaults}
\begin{figure}[t]
	\begin{lstlisting}[
basicstyle=\ttfamily\scriptsize,
texcl,
emph={[2]send_buf, send_counts},
emphstyle={[2]\color{gray}},
escapebegin=\;\color{my-dark-red},
emph={[3]vector, std, exclusive_scan},
emphstyle={[3]\color{my-purple}}
]
std::vector<T> v = ...; // fill with data

int size, rank;
MPI_Comm_size(comm, &size);
MPI_Comm_rank(comm, &rank);
std::vector<int> rc(size), rd(size);
rc[rank] = v.size();
// exchange counts
MPI_Allgather(MPI_IN_PLACE, 0, MPI_DATATYPE_NULL,
 rc.data(), 1, MPI_INT, comm);
// compute displacements
std::exclusive_scan(rc.begin(), rc.end(), rd.begin(), 0);
int n_glob = rc.back() + rd.back();
// allocate receive buffer
std::vector<T> v_glob(n_glob);
// exchange
MPI_Allgatherv(v.data(), v.size(), MPI_TYPE, v_glob.data(),
 rc.data(), rd.data(), MPI_TYPE, comm);
\end{lstlisting}
	\caption{
          Allgathering of a vector using MPI.}
	\label{fig:kamping_vs_mpi}
\end{figure}

As discussed previously, MPI calls often allow for computing useful default values for an operation based on only a small subset of parameters.
As an example, consider the case, where we want to perform an \MPI{allgatherv}, where each rank initially holds an \lstinline!std::vector! of varying size and we want to concatenate them to a global vector on each rank.
The send count and data type can be directly inferred from the vector's size and \lstinline!value_type! (see \cref{sec:types} for more details on automatic type deduction).
Receive counts and displacements can be computed by an \MPI{allgather} of all send counts followed by an exclusive prefix sum over them (see \cref{fig:kamping_vs_mpi}).
While this is a common pattern, none of the other \Cpp bindings allows avoiding all this boilerplate code.
Boost.MPI offers various overloaded functions that allow the user to omit explicit displacements, but the counts have to be communicated.
RWTH-MPI does provide an overload that gathers the counts internally, but it only works with \lstinline!MPI_IN_PLACE!, which requires the send data to be already provided at the correct position on each rank. To achieve this the user has to manually exchange count information upfront%
\footnote{Full example code can be found at \url{https://github.com/kamping-site/kamping-examples/tree/main/include/vector_allgather/}}.
This leaves us with a situation where the usability of \Cpp bindings depends on whether the implementers had this particular use case in mind and provided a default option for it.

To address this problem, we choose an alternative approach inspired by \emph{named parameters}, where parameters passed to a function can be named at the caller site and passed in arbitrary order (as known from languages like Python).
Internally, named parameters are realized as \emph{factory functions}~\cite{Gamma1995} which construct lightweight parameter objects encapsulating the parameter type (i.e., send buffer, send counts, $\dots$) and the corresponding data.
This allows us to check for the presence of each parameter at compile time and to compute default values only if the respective parameter is omitted, without resorting to many overloads exploring the complete combinatorial explosion of parameters.
To avoid runtime overhead, we rely on template metaprogramming to only generate the code paths required for computing missing parameters at compile time.
Gathering a vector in \kamping then becomes a one-liner as shown in \cref{fig:kamping-intro}.
The implementations using other bindings are more verbose (see \cref{tab:loc}).

The flexibility of the named parameter interface allows users to iteratively adapt their existing code to \kamping, as shown in \cref{fig:kamping-step-by-step}.

\begin{figure}[b]
  \begin{lstlisting}[
    basicstyle=\ttfamily\scriptsize,
    emph={[2]send_buf, send_counts, recv_buf, recv_counts, recv_displs, send_recv_buf},
    emphstyle={[2]\color{gray}},
    emph={[3]vector, std, exclusive_scan},
    emphstyle={[3]\color{my-purple}}
    ]
// Version 1: using KaMPIng's interface
std::vector<int> rc(comm.size()), rd(comm.size());
rc[comm.rank()] = v.size();
comm.allgather(send_recv_buf(rc));
std::exclusive_scan(rc.begin(), rc.end(), rd.begin(), 0);
std::vector<T> v_glob(rc.back() + rd.back());
comm.allgatherv(send_buf(v), recv_buf(v_glob),
 recv_counts(rc), recv_displs(rd));

// Version 2: displacements are computed implicitly
std::vector<int> rc(comm.size());
rc[comm.rank()] = v.size();
comm.allgather(send_recv_buf(rc));
std::vector<T> v_glob;
comm.allgatherv(send_buf(v),
 recv_buf<resize_to_fit>(v_glob), recv_counts(rc));

// Version 3: counts are automatically exchanged
// and result is returned by value
std::vector<T> v_glob = comm.allgatherv(send_buf(v));
\end{lstlisting}
\caption{The vector allgather MPI example from \cref{fig:kamping_vs_mpi} can be gradually migrated to use more of \kamping's features.}
\label{fig:kamping-step-by-step}
\end{figure}

\subsection{Input and Output Parameters}
\label{sec:in-out-parameters}
\kamping extends MPI's definition of \emph{in(put)}- and \emph{out(put)}-parameters.
With an in-parameter, the caller provides data to the wrapped MPI call, such as with \lstinline!send_buf(data)!.
By passing an out-parameter, e.g., \lstinline!recv_counts_out()!, the caller asks the library to compute the requested parameter and return its result.
Most MPI parameters like send displacements, receive counts, etc. can either be passed as in- or out-parameters since they can be internally deduced in many cases using additional computation or communication as outlined in \cref{sec:defaults}.

The parameter type is determined by the named parameter factory functions:
To give an example, \lstinline!send_displs(data)! creates an in-parameter containing the send displacements as specified in \lstinline!data! whereas \lstinline!send_displs_out()! creates an out-parameter signaling to return the send displacements by value.

Since one is primarily interested in the receive buffer in MPI calls, this parameter is always implicitly returned by \kamping.
To retrieve other parameters from the wrapped MPI call one has to explicitly pass the corresponding out-parameter.
This is a major improvement over previous MPI libraries, which simply mimic MPI's C-Interface and return output data next to the receive buffer by pointer or reference.
This is not in line with the \Cpp core guidelines which strongly suggest returning output data by value \cite[F.20]{CppCoreGuidelines}.
Furthermore, it is often unclear which additional parameters are computed by the library, as the overloaded wrapped MPI function calls differ only in the number of function arguments.
Combined with the large number of parameters of MPI calls, it is hard to see which argument refers to which parameter when looking at the code.

\kamping on the other hand improves this situation in two regards:
\begin{enumerate}
	\item The caller can decide which non-required parameters they want \kamping to compute internally.
	      By the named parameter approach this decision is clearly documented in the source code and correctness does not depend on a common understanding of the parameter order.
\item For each requested out-parameter, the caller can individually decide how the data is returned.
\end{enumerate}
In the following, this is illustrated with a call to the wrapped \MPI{Allgatherv} function.

\begin{lstlisting}[
  basicstyle=\ttfamily\scriptsize,
  emph={MPI_Allgatherv, MPI_Alltoallv, MPI_Allgather, alltoallv, allgather, allgatherv, Communicator, MPI_Comm_size},
  emphstyle={\bfseries\color{blue}},
  emph={[2]send_buf, send_counts, recv_counts_out},
  emphstyle={[2]\color{gray}},
  emph={[3]vector, std, exclusive_scan},
  emphstyle={[3]\color{my-purple}}
]
auto result = comm.allgatherv(send_buf(v),
                              recv_counts_out());
auto recv_buf = result.extract_recv_buf();
auto counts = result.extract_recv_counts();
\end{lstlisting}

The above call to \lstinline!allgatherv! returns a result object containing the (implicitly) requested receive buffer and the receive counts.
These can then be extracted from the result object using \emph{move} semantics.
It is furthermore possible to directly decompose the result object using \Cpp's \emph{structured bindings} which simplifies the call to

\begin{lstlisting}[
  basicstyle=\ttfamily\scriptsize,
  emph={MPI_Allgatherv, MPI_Alltoallv, MPI_Allgather, alltoallv, allgather, allgatherv, Communicator, MPI_Comm_size},
  emphstyle={\bfseries\color{blue}},
  emph={[2]send_buf, send_counts, recv_counts_out},
  emphstyle={[2]\color{gray}},
  emph={[3]vector, std, exclusive_scan},
  emphstyle={[3]\color{my-purple}}
]
auto [recv_buf, counts] = comm.allgatherv(send_buf(v),
                              recv_counts_out());
\end{lstlisting}

A shortcoming of returning by value is the redundant memory allocation in case a previously allocated container could be reused instead.
In \kamping, we offer two solutions for this scenario.
For containers supporting \Cpp move semantics, a previously user-allocated container can be simply moved to the underlying call and is then subsequently returned with the result object by value.
\begin{lstlisting}[
  basicstyle=\ttfamily\scriptsize,
  emph={MPI_Allgatherv, MPI_Alltoallv, MPI_Allgather, alltoallv, allgather, allgatherv, Communicator, MPI_Comm_size},
  emphstyle={\bfseries\color{blue}},
  emph={[2]send_buf, recv_buf, send_counts, recv_counts_out},
  emphstyle={[2]\color{gray}},
  emph={[3]vector, std, exclusive_scan, move},
  emphstyle={[3]\color{my-purple}}
]
std::vector<T> tmp = ...;
// tmp is moved to the underlying call where the
// storage is reused for the recv buffer
auto recv_buffer = comm.allgatherv(
                          send_buf(v),
                          recv_buf(std::move(tmp)));

\end{lstlisting}

If there is no efficient way to support move semantics for a container type, it is also possible to pass the container via reference to the underlying call.
The data computed by \kamping for the requested parameter will then be written directly to the specified memory location.

\begin{lstlisting}[
  basicstyle=\ttfamily\scriptsize,
  emph={MPI_Allgatherv, MPI_Alltoallv, MPI_Allgather, alltoallv, allgather, allgatherv, Communicator, MPI_Comm_size},
  emphstyle={\bfseries\color{blue}},
  emph={[2]send_buf, recv_buf, send_counts, recv_counts_out},
  emphstyle={[2]\color{gray}},
  emph={[3]vector, std, exclusive_scan},
  emphstyle={[3]\color{my-purple}}
]
std::vector<T> recv_buffer = ...;
// data is written into recv_buffer directly
comm.allgatherv(send_buf(v),
                recv_buf(recv_buffer));
\end{lstlisting}

\subsection{Controlling Memory Allocation}
\label{sec:memory}
Previous MPI wrappers have no unified way of controlling memory allocation.
They either accept containers that are always resized to fit, or, if resizing is not desired, the user has to pass raw pointers directly.
They also offer no control over allocation happening for default parameter computation.

\kamping allows for fine-grained control over memory management.
Each (out-)parameter accepting a container takes an optional template parameter indicating its \emph{resize policy}, which controls whether it is always resized to fit, resized if it does not have enough space to store the result, or performs no checking and assumes that the capacity of the container is large enough, which is the default.

\begin{lstlisting}[
  basicstyle=\ttfamily\scriptsize,
  emph={MPI_Allgatherv, MPI_Alltoallv, MPI_Allgather, alltoallv, allgather, allgatherv, Communicator, MPI_Comm_size},
  emphstyle={\bfseries\color{blue}},
  emph={[2]send_buf, recv_buf, send_counts, recv_counts_out},
  emphstyle={[2]\color{gray}},
  emph={[3]vector, std, exclusive_scan},
  emphstyle={[3]\color{my-purple}}
]
std::vector<T> recv_buffer; // has to be resized
std::vector<int> counts(comm.size()); // size large
                                      // enough
comm.allgatherv(send_buf(v),
    recv_buf<resize_to_fit>(recv_buffer),
    recv_counts_out(counts));
\end{lstlisting}

If \kamping has to create auxiliary data structures to compute missing parameters, the user may either pass preallocated containers to use or provide the container's type via template parameters.
Recall that additional allocation is omitted entirely when parameters are provided by the user.

As all of this relies on template metaprogramming, there is no additional overhead compared to a hand-rolled implementation.
This flexibility allows to quickly implement distributed algorithms and then iteratively fine-tune memory allocations and library inferred values.
This facilitates an algorithm engineering workflow that involves iterative refinement of implementations and analysis through experimentation.

\subsection{Using Custom Types}
\label{sec:types}
HPC applications use a variety of data types that need to be communicated.
Beyond \emph{basic data types} corresponding directly to \Cpp{}'s built-in types,
MPI allows for complex \emph{derived data types} using type constructors, such as \MPI{Type\_create\_struct} and \MPI{Type\_create\_contiguous}.

C's lack of type introspection forces users to always pass the type explicitly to a communication call, which is both tedious and error-prone, as type definitions need to be kept in sync with the actual data layout.

Template-metaprogramming enables mapping arbitrary \Cpp types one-to-one to MPI data types (we call this a \emph{static type}).
MPI's derived data types, however, form a superset of \Cpp data types.
This is because MPI allows constructing arbitrary type signatures with sizes known only at runtime (we call this a \emph{dynamic type}).
\kamping provides support for static and dynamic types and offers implicit static type construction without performance pitfalls.
Sometimes, applications need to communicate unstructured and complex data types off the critical code path.
To support this with minimal code overhead, \kamping provides transparent serialization support.
In the following, we discuss these aspects in more detail.

\subsubsection{Static derived data types at compile time}

\kamping maps basic \Cpp data types to their MPI counterparts and supports complex data types on homogeneous systems if they are \emph{trivially copyable}, i.e., the \Cpp standard guarantees that they can be copied into a \lstinline{char} array.
In this case, we create a contiguous type using \MPI{Type\_contiguous} with the appropriate number of bytes, as this provides a sensible default which usually achieves better performance than using \MPI{Type\_struct} (see \cref{sec:types-experiment}).
For all other types, the user can directly provide static type definitions by providing an explicit instantiation of the \lstinline!mpi_type_traits!
template for the desired type, which describes how to construct a matching \lstinline!MPI_Datatype!, as show in \cref{fig:custom_types}.
While this allows for building data types using MPI's type constructors, constructing a correct type for a given \Cpp struct is error-prone as the programmer has to keep the type-construction calls in sync with the data type.
We leverage the \emph{PFR} library~\cite{Polukhin2016} to automatically generate MPI type definitions for user-provided structs at compile time.
This can be enabled by inheriting from the type constructor when defining the type trait (see \cref{fig:custom_types}).
\begin{figure}
\begin{lstlisting}[
    basicstyle=\ttfamily\scriptsize,
    morekeywords={constexpr, size_t},
    emph={[3]vector, std, array},
    emphstyle={[3]\color{my-purple}},
    emph={mpi_type_traits, struct_type, data_type},
    emphstyle={\bfseries\color{blue}},
]
struct MyType {
   int a;
   double b;
   char c;
   std::array<int, 3> d;
};
namespace kamping {
// using KaMPIng's built-in struct serializer
template <>
struct mpi_type_traits<MyType> : struct_type<MyType> {};

// or using an explicitly constructed type
template <>
struct mpi_type_traits<MyType> {
   static constexpr bool has_to_be_committed = true;
   static MPI_Datatype data_type() {
       MPI_Datatype type;
       MPI_Type_create_*(..., &type);
       return type;
   }
};
} // namespace kamping
\end{lstlisting}
    \caption{Defining custom static types using automatic type reflection or a custom type definition.}
    \label{fig:custom_types}
\end{figure}

MPI requires the user to initialize and deallocate non-built-in types.
\kamping archives both transparently to the user by exploiting the Construct-On-First-Use-Idiom \footnote{\url{https://isocpp.org/wiki/faq/ctors\#static-init-order-on-first-use}}.

Existing MPI bindings also support non-built-in static types to some extent.
For Boost.MPI that is dependent on the definition of a serialization function.
Similar to \kamping, types are managed using a global type pool, but each operation incurs a runtime type lookup.
MPL and RWTH-MPI use the Construct-On-First-Use-Idiom to commit types before first use, and MPL also uses PFR to provide automatic type definitions by using reflection.
Opposed to \kamping, types are not properly freed, which may result in resource leakage.
RWTH-MPI allows custom static type definitions, but the user is responsible for committing and freeing types.

\subsubsection{Support for dynamic types}
For data types constructed at runtime, \kamping supports passing explicit types to operations directly, by providing optional type parameters.

In \kamping, dynamic types can currently only be constructed using MPI's type constructors.
MPL on the other hand provides a runtime type interface mirroring MPI's type constructors using the builder pattern, which encapsulates the constructed types in so-called \emph{layouts}.
While MPL's collective operations are tightly interleaved with the layout system, which results in verbose code, its type construction is a powerful feature that we plan to integrate as the default way for constructing dynamic types in \kamping{}.
RWTH-MPI offers no support for dynamic types.

\subsubsection{Communicating arbitrary data using serialization}
Some applications require communicating non-contiguous data which is (partially) located on the heap, e.g., \lstinline{std::string} or \lstinline{std::unordered_map} and cannot be represented using MPI datatypes.
These have to be packed into a contiguous buffer before communication.
\kamping facilitates this by providing serialization support, which is both highly tuneable and transparent to the user.
We rely on the popular \Cpp serialization library \emph{Cereal}~\cite{Grant2017}, which supports STL data types and allows providing serialization routines for custom types.
While serialization is transparent to the user, i.e., the user never sees the serialized data, it still has to be explicitly enabled as it usually incurs hidden costs for allocating memory for serialized data and performing \mbox{(de-)serialization}.
Through Cereal's flexible design, serialization in \kamping is also highly configurable; allowing users to specify custom serialization functions and archives, e.g., binary formats, JSON, or XML.
See \cref{fig:serialization} for an example of how to use serialization in \kamping.

\begin{figure}
    \begin{lstlisting}[
  basicstyle=\ttfamily\scriptsize,
  emph={send, recv},
  emphstyle={\bfseries\color{blue}},
  emph={[2]send_buf, send_counts, recv_counts_out, recv_buf},
  emphstyle={[2]\color{gray}},
  emph={[3]vector, std, exclusive_scan},
  emphstyle={[3]\color{my-purple}},
  emph={[4]as_serialized, as_deserializable},
  emphstyle={[4]\color{my-teal}},
]
using dict = std::unordered_map<std::string, std::string>;
dict data = ...;
comm.send(send_buf(as_serialized(data)), ...);

dict recv_dict = comm.recv(
  recv_buf(as_deserializable<dict>())
);
\end{lstlisting}
    \caption{Usage of \kamping's serialization.}
    \label{fig:serialization}
\end{figure}

Besides \kamping, Boost.MPI is the only library offering serialization support, but is tightly coupled with other Boost libraries and serialization is performed implicitly; if a type is not marked as an MPI data type, serialization is used.
This makes it hard to see whether costly serialization is involved by just looking at the code.

We are convinced, that using serialization implicitly should be avoided by zero-overhead MPI bindings, as hidden serialization incurs hidden runtime and memory overheads.

\subsubsection{Towards sensible defaults for type construction}
\label{sec:types-experiment}

By default, \kamping maps trivially copyable types to a type interpreted as a contiguous sequence of bytes.
When defining a struct type where the members have alignment gaps, MPI does not include the gap in the communicated data.
This requires non-contiguous memory accesses, which may be slower than copying whole memory blocks to the communication hardware (MPI standard~\cite[\S 5.1.6]{Forum2023}).
The standard suggests introducing dummy struct members to fill these gaps, but this requires the user to modify their non-MPI code.
By using a type consisting of contiguous bytes when valid with respect to the \Cpp standard, we enable this more efficient default transparent to the user.
Preliminary experiments also confirm this in practice and further highlight that serialization incurs a non-negligible overhead, which is the reason why \kamping uses serialization only if explicitly enabled.

\subsection{Enabling Safety for Non-blocking Communication}
\label{sec:nonblocking}
Non-blocking communication in MPI is important for both correct and performant applications.
MPI allows to \emph{initiate} an operation, which returns a \emph{request handle} to the user.
A user then has to \emph{complete} the request, by either testing for completion of the request using \MPI{test} or using \MPI{wait} to block until the request is completed.
The semantics of MPI only allow updating send buffers or reading from receive buffers taking part in a previously initiated operation when the corresponding request has been completed.
This introduces an additional source for programming errors, as MPI does not hinder users from accessing the memory locations regardless of completion status.

For asynchronous (I/O) operations the \Cpp standard library provides \lstinline{std::future}, which allows querying or waiting for the result of an asynchronous operation, only returning a value once the operation has completed.
Using \lstinline{std::future} to provide a safe interface for non-blocking communicating is not possible, as they are tied to asynchronous progress happening in the background, which the MPI standard does not guarantee.

To solve this, we introduce a similar concept called a \emph{non-blocking MPI result}, which encapsulates an \MPI{request} and the data returned by value from the operation, as described in \cref{sec:in-out-parameters}.
The data is only returned to the user by calling \lstinline{result.wait()} which internally completes the request.
Calling \lstinline{result.test()} returns an \lstinline{std::optional}, which only contains the returned data if the request is completed, and otherwise returns \lstinline{std::nullopt}.
This ensures that a user can only access valid received data.

To also prevent unwanted access to send buffers during non-blocking operations, the user can move the data into the call.
The non-blocking result then assures that the data lives long enough, and it is re-returned to the user upon completion of the call.
This happens without any copying of data by relying on \Cpp's move semantics.
See \cref{fig:nonblocking-safety} for an example.

\begin{figure}
	\begin{lstlisting}[
  basicstyle=\ttfamily\scriptsize,
  emph={isend, irecv, wait},
  emphstyle={\bfseries\color{blue}},
  emph={[2]send_buf_out, destination, recv_count},
  emphstyle={[2]\color{gray}},
  emph={[3]vector, std, exclusive_scan, move},
  emphstyle={[3]\color{my-purple}}
]
std::vector<int> v = ...;
auto r1 = comm.isend(
    send_buf_out(std::move(v)), destination(1)
);

v = r1.wait(); // v is moved back to caller after
               // request is complete

auto r2 = comm.irecv<int>(recv_count(42));
std::vector<int> data = r2.wait(); // data only returned
                                   // after request
                                   // is complete
\end{lstlisting}
	\caption{Example of non-blocking safety in \kamping.}
	\label{fig:nonblocking-safety}
\end{figure}

This is enabled by \kamping's distinct parameter handling, making it the first \Cpp MPI library that provides safety guarantees for non-blocking communication.
Opposed to that, other MPI bindings offer no enhanced safety features for the data involved in a non-blocking call, but only return request handles.
Here the user is still responsible for ensuring that no invalid data access happens while an operation is in progress.
Only \emph{rsmpi} provides similar guarantees, powered by Rust's ownership model.

Another feature \kamping provides to facilitate working with non-blocking MPI are \emph{request pools}, which allow for easy completion of multiple requests.
The user only needs to submit the request associated with the call to such a pool.
While the current implementation just collects them in an unbounded array, requests pool are designed with extensibility in mind, to enable more sophisticated variants.
For example, we are currently working on a request pool with a fixed number of slots, internally maintaining free slots, which allows limiting the number of concurrent non-blocking requests.

\subsection{Extensibility}
\label{sec:expandability}
While the main goal of \kamping is to design \Cpp bindings for MPI which can be used in any MPI application, designing a one-size-fits-all library is hard.
To enable further flexibility we therefore
designed it with extensibility and compatibility with existing MPI code in mind, to allow easy extension and alteration of its core features.
On the one hand, an existing code base can be gradually migrated to use only some parts of \kamping, because it is fully compatible with native MPI objects, such as request or type handles.
On the other hand, \kamping's plugin interface allows overriding member functions of a communicator object (e.g., collectives) and adding additional functionality without changing existing application code.
The library allows plugin implementers to define new named parameters to enable the full named parameter flexibility also for these library extensions.
This architecture allows us to keep \kamping's core library small while providing a base for third-party general-purpose MPI libraries and keeping maintenance low, in order not to follow the fate of the official MPI \Cpp interface.

With \kamping, we already ship multiple plugins extending the functionality of the current MPI standard (see \cref{sec:general-building-blocks}).

\subsection{More Safety Features for MPI}
\label{sec:more-safety}
In addition to the ease of use and safety features introduced by \kamping's parameter handling, it also provides other means of preventing sources of errors in MPI code.
\paragraph*{Error handling}
MPI notifies users of errors by return codes.
Here, MPI makes no distinction between failures, such as insufficient buffer space or node failures, which may be recoverable, and usage errors, such as providing invalid parameters.

\kamping tries to improve upon this by using three major techniques, following the \Cpp core guidelines: using exceptions for failures~\cite[E.2]{CppCoreGuidelines}, catching usage errors at compile time whenever possible~\cite[P.5]{CppCoreGuidelines}, and making heavy use of assertions at runtime.
While \Cpp's template metaprogramming is notorious for complex and hard-to-read compiler errors, we try to ensure that compile-time assertions fail early and provide helpful human-readable error messages, e.g., for missing parameters or incompatible types.
For example, when the user does not provide a required parameter to a collective operation, the error message indicates which parameter is missing during compile time.
\kamping also includes many runtime assertions verifying MPI invariants, that are grouped in different levels, ranging from lightweight checks to assertions involving additional communication.
The assertions can be completely disabled level-by-level.
The use of exceptions can also be completely disabled, and \kamping allows overriding the error handling strategy using the plugin system.

In contrast to that, other MPI bindings either always convert MPI errors to exceptions or do not provide any error handling.

\paragraph*{Simplified \lstinline{MPI_IN_PLACE}}
\kamping also tries to prevent programming errors when working with in-place MPI operations.
Consider using the in-place variant of \MPI{allgather}, as in \cref{fig:kamping_vs_mpi}:
Using MPI's C interface, the user has to explicitly pass \lstinline{MPI_IN_PLACE} on all ranks as send buffer and the send count and type parameters are ignored.
\kamping's concept of in-out-parameters simplifies the call semantics of in-place calls.
If a user passes data as \lstinline{send_recv_buf} instead of \lstinline{send_buf}, then \kamping automatically passes the correct arguments to the underlying in-place call and issues a compilation error if the user provides an argument which would be ignored by the in-place call.
This is also compatible with move semantics, allowing for concise in-place calls as shown below:

\begin{lstlisting}[
  basicstyle=\ttfamily\scriptsize,
  emph={MPI_Allgatherv, MPI_Alltoallv, MPI_Allgather, alltoallv, allgather, allgatherv, Communicator, MPI_Comm_size},
  emphstyle={\bfseries\color{blue}},
  emph={[2]send_buf, recv_buf, send_counts, recv_counts_out, send_recv_buf},
  emphstyle={[2]\color{gray}},
  emph={[3]vector, std, exclusive_scan, move},
  emphstyle={[3]\color{my-purple}}
]
std::vector<int> data(comm.size());
data[comm.rank()] = ...;
data = comm.allgather(send_recv_buf(std::move(data)));
\end{lstlisting}

\subsection{Implementation Details}
\label{sec:implementation-details}
\kamping makes extensive use of \Cpp's template metaprogramming capabilities. For example, using \lstinline{constexpr if} and \emph{SFINAE}~\cite[8.4]{Vandevoorde2018} (Substitution Failure is Not an Error), often allows us to eliminate runtime control flow constructs in the wrappers around MPI calls, preventing costly branch mispredictions.
Named parameters are implemented using template parameter packs of parameter objects.
Compile-time checks for the absence of each parameter then allow to conditionally enable logic for computing default values via \lstinline{constexpr if}.

All containers passed as arguments are wrapped in a templated \emph{DataBuffer} which handles ownership and modifiability transparently at compile time\footnote{Because this works with any STL-compliant container, KaMPIng also supports accelerator-aware MPI implementations directly. Pointers or containers (like \lstinline[basicstyle=\footnotesize\ttfamily]{thrust::device_vector}) to device memory can be passed just like normal containers.}.
These data buffers are moved to a templated result object, which is returned to the user and can be destructed using structured bindings as described in \cref{sec:in-out-parameters}.
As we do not copy but move the data, this imposes nearly zero overhead.

All wrapped MPI functionality has been extensively tested using a large number of parameter combinations as part of our development cycle.
All (unit-)tests are executed using the C++ compilers clang and gcc (in debug and release mode) and OpenMPI.
We use MPI's profiling interface to ensure that only the expected MPI calls are issued if \kamping calls MPI internally to compute default values.
Additionally, we evaluated our benchmarks using OpenMPI and IntelMPI on the supercomputers SuperMUC-NG and HoreKa, as detailed in the following section.

\section{Integrating \kamping into Real-World Applications}
\label{sec:evaluation}

To highlight the usability of our library, we have integrated \kamping into multiple
(research) applications, ranging from sorting (sample sort and suffix sorting) over graph algorithms (BFS and label propagation) to a large phylogenetic interference tool.
Experiments backing our (near) zero overhead claim are executed on up to 256 compute nodes of SuperMUC-NG%
\footnote{%
  We also conducted some experiments on the smaller HoreKa supercomputer.
  As the results obtained there are similar to our findings from SuperMUC-NG, we omit them here.}%
, where each node is equipped with an Intel Skylake Xeon Platinum 8174 processor with \num{48} cores.
The internal interconnect is a fast OmniPath network with \SI{100}{\giga\bit\per\second}.
Our code is compiled with g++-12.2.0 and Intel MPI 2021 using optimization level \texttt{-O3}.

\begin{table}
  \caption{Lines of code for examples using \kamping vs.\ other bindings\protect\footnotemark.}
  \label{tab:loc}
    \begin{tabularx}{\linewidth}{lccccc}
    \toprule
                     & MPI & Boost.MPI & RWTH-MPI & MPL & \textbf{\kamping} \\
    \midrule
    vector allgather & 14  & 5     & 5    & 12  & 1                 \\
    sample sort      & 32  & 30    & 21   & 37  & 16                \\
    BFS              & 46  & 42    & 32   & 49  & 22                \\
    \bottomrule
  \end{tabularx}

\end{table}
\footnotetext{The source codes for all three examples are available at \url{https://github.com/kamping-site/kamping-examples/}.}

\subsection{Sorting}
\label{sec:sorting}
As a first example, we use a textbook distributed sample sort~\cite{Sanders2019} with \cref{code:sample-sort-kamping} showing a prototypical implementation in \kamping.
We have implemented the sample sort algorithm using all previously discussed \Cpp MPI bindings comparably, where all shared parts of the code have been extracted to functions and the code has been formatted identically with \texttt{clang-format} using the \texttt{Google} style template.
In this setting, we require only $16$ lines of code (LOC) using \kamping while the plain MPI example requires $32$ LOC.
The lines of code for all bindings can be found in \cref{tab:loc}.
\Cref{fig:kamping-performance} shows the running time of the different implementations in a weak-scaling experiment.
We sort a distributed array with $10^{6}$ 64-bit integers per rank, which are drawn uniformly at random.

We see that \kamping introduces no additional overhead compared to a hand-rolled implementation in plain MPI or other libraries, but makes the implementation a lot easier to read and write while being more flexible.

\paragraph*{Suffix Array Construction}
For a more complex example, we consider an application from text processing:
We sort all suffixes of a text lexicographically, i.e., we compute the suffix array~\cite{Manber1993}.
Here, we implemented two algorithms: DCX~\cite{Karkkainen2006} and Prefix Doubling~\cite{Manber1993}.
Our \kamping implementation of DCX requires 1\,264~LOC whereas the plain MPI implementation~\cite{Bingmann2018pDCX} needs 1\,396~LOC.
The additional 9.5\,\% of code is mostly due to boilerplate code, e.g., distributing send counts for \MPI{Alltoallv} and the tedious construction of MPI types.
Likewise, our \kamping implementation of the Prefix Doubling algorithm requires 163~LOC.
An existing plain MPI implementation of the same algorithm~\cite{Fischer2019} needs 426~LOC (not counting the 1\,442~LOC for wrapped MPI functionality used by the plain MPI implementation).
Even when using a high-level distributed programming framework, implementing the Prefix Doubling algorithm still requires 266~LOC~\cite{Bingmann2018}.

\begin{figure}
\begin{lstlisting}[
		basicstyle=\ttfamily\scriptsize,
		emphstyle={[2]\color{gray}},
    emph={[2]send_buf, send_counts},
		emph={[3]vector, sort, sample, std, mt19937, random_device},
		emphstyle={[3]\color{my-purple}},
		emph={[4]pick_splitters, build_buckets},
		emphstyle={[4]\color{my-teal}},
		firstline=4
	]
template<typename T>
void sort(std::vector<T>& data, MPI_Comm comm_) {
  using namespace std;
  Communicator comm(comm_);
  const size_t num_samples = 16 * log2(comm.size()) + 1;
  vector<T> lsamples(num_samples);
  sample(data.begin(), data.end(), lsamples.begin(),
        num_samples, mt19937{random_device{}()});
  auto gsamples = comm.allgather(send_buf(lsamples));
  sort(gsamples.begin(), gsamples.end());
  for (size_t i = 0; i < comm.size() - 1; i++) {
    gsamples[i] = gsamples[num_samples * (i + 1)];
  }
  gsamples.resize(comm.size() - 1);
  vector<vector<T>> buckets = build_buckets(data, gsamples);
  data.clear();
  vector<int> scounts;
  for (auto &bucket : buckets) {
    data.insert(data.end(), bucket.begin(), bucket.end());
    scounts.push_back(bucket.size());
  }
  data = comm.alltoallv(
    send_buf(data), send_counts(scounts)
  );
  sort(data.begin(), data.end());
}
\end{lstlisting}
	\caption{Distributed sample sort using \kamping.}
	\label{code:sample-sort-kamping}
\end{figure}

\pgfplotsset{
	major grid style={thin,dotted},
	minor grid style={thin,dotted},
	ymajorgrids,
	yminorgrids,
	every axis/.append style={
			line width=0.7pt,
			tick style={
					line cap=round,
					thin,
					major tick length=4pt,
					minor tick length=2pt,
				},
		},
	legend style={
			line width=0.25pt,
			/tikz/every even column/.append style={column sep=3mm,black},
			/tikz/every odd column/.append style={black},
		},
	legend style={font=\small},
	enlarge x limits=0.04,
	every tick label/.append style={font=\footnotesize},
	every axis label/.append style={font=\small},
	every axis y label/.append style={yshift=-1ex},
	axis lines*=left,
	xlabel near ticks,
	ylabel near ticks,
	axis lines*=left,
	label style={font=\footnotesize},
	tick label style={font=\footnotesize},
	MPI/.style={
			color=my-red,
			dashed,
			mark=square,
			mark options={solid},
		},
	Boost/.style={
			color=my-purple,
			dashed,
			mark=diamond,
			mark options={solid},
		},
	mpl/.style={
			color=my-indigo,
			dashed,
			mark=triangle,
			mark options={solid},
		},
	KaMPIng/.style={
			color=my-orange,
			dashed,
			mark=x,
			mark options={solid},
		},
	RWTH/.style={
			color=my-teal,
			dashed,
			mark=o,
			mark options={solid},
		},
}

\begin{figure}
  \centering
  \input{plots/sorting.pgf}
  \caption{Running time of sample sort using different MPI bindings.}
  \label{fig:kamping-performance}
\end{figure}

\subsection{Graph Algorithms}
Currently, most state-of-the-art HPC platforms are mainly designed for numerical applications with fairly regular data access and communication patterns.
However, data-intensive irregular workloads become more and more frequent, for example, in materials science or data analysis tasks such as (human) brain analysis.
Graphs are commonly used to represent data sets in these applications and we therefore require efficient distributed graph algorithms either for network analysis or as important building blocks for more complex applications \cite{Slota2016, davis1985evolution,hotzer2015phase}.
To demonstrate the applicability of \kamping in this setting, we provide a simple distributed breadth-first search (BFS) implementation in \cref{code:bfs-kamping}.
We assume the graph to be distributed among the ranks with each rank holding a subset of the vertices and their incident edges.
Locally, the graph is represented as an adjacency array.
For each vertex $v$, the \texttt{bfs} returns the distance (number of hops) between the source vertex \texttt{s} and $v$.
\kamping's utility function \lstinline!with_flattened(...)!, which \emph{flattens} a container of nested destination-message pairs by transforming it into a contiguous range while also providing send counts, proves to be especially useful.

As for the sample sort example, we implemented the distributed BFS algorithm using all previously discussed (\Cpp) MPI bindings comparably.
The implementations only differ for the frontier exchange and completion logic, which can be implemented in \kamping using only $22$ lines of code, whereas plain MPI requires $46$ lines.
Our closest competitor regarding code length is RWTH-MPI with $32$ LOC (see \cref{tab:loc} for all other bindings) \footnote{The code considered here is structured slightly differently compared to \cref{code:bfs-kamping}, which has been shortened for readability. See \url{https://github.com/kamping-site/kamping-examples/tree/main/include/bfs/} for full implementations.}.

In \cref{plot:bfs-benchmark} we evaluate these implementations using a variety of different graph families and observe that \kamping introduces no additional overhead compared to plain MPI.
\kamping also provides optimized collective operations, which provide better scalability than \MPI{alltoallv} on some graph families, which we discuss in \cref{sec:sparse-alltoall}.

Our base \kamping implementation as well as RWTH-MPI and Boost.MPI (omitted in the plot) are always on par with the MPI implementation.
MPL (also omitted in the plot) internally uses \MPI{Alltoallw} for \alltoall{} exchanges and is (considerably) slower than MPI on all configurations, as already discussed previously~\cite{Ghosh2021}.

\begin{figure}
\begin{lstlisting}[
  basicstyle=\ttfamily\scriptsize,
  emphstyle={[2]\color{gray}},
  emph={[3]vector, sort, sample, std, mt19937, random_device, exclusive_scan, log2, unordered_map, move, numeric_limits},
  emphstyle={[3]\color{my-purple}},
  emph={[4]bfs, is_empty, exchange, expand_frontier},
  emphstyle={[4]\color{my-teal}},
  morekeywords={fn, let, mut, where, as, constexpr, size_t},
  firstline=5,
  lastline=35
  ]
using VId = size_t;
using VBuf = std::vector<VId>;
constexpr VId undef = std::numeric_limits<VId>::max();

bool is_empty(auto &frontier, Communicator const& comm) {
  return comm.allreduce_single(send_buf(frontier.empty()),
                               op(std::logical_and<>{}));
}
VBuf exchange(auto frontier, Communicator const &comm) {
  return with_flattened(frontier, comm.size()).call(
    [&](auto... flattened) {
      return comm.alltoallv(std::move(flattened)...);
    });
}
vector<size_t> bfs(Graph const &g, VId s, MPI_Comm _comm) {
  Communicator comm(_comm);
  VBuf frontier;
  std::unordered_map<int, VBuf> next_frontier;
  if (g.is_local(s)) {
    frontier.push_back(s);
  }
  std::vector<size_t> dist(g.local_size(), undef);
  size_t level = 0;
  while(!is_empty(frontier, comm)) {
    next_frontier = expand_frontier(g, frontier, dist, level);
    frontier = exchange(std::move(next_frontier), comm);
    next_frontier.clear();
    ++level;
  }
  return dist;
}
\end{lstlisting}
  \caption{Distributed BFS using \kamping.}
  \label{code:bfs-kamping}
\end{figure}

\begin{figure}[!ht]
  \input{plots/bfs_running_times.pgf}
  \caption{Running time of BFS comparing MPI and \kamping and additional \alltoall{} implementations.}
  \label{plot:bfs-benchmark}
\end{figure}


\paragraph*{Graph Partitioning}
As a more complex showcase, we integrated \kamping into the state-of-the-art distributed multilevel graph partitioner dKaMinPar~\cite{dKaMinPar}, consisting of roughly 30\,000 LOC and including its own abstraction layer with specialized graph-specific communication primitives over plain MPI.
The partitioner uses size-constrained label propagation to iteratively cluster and contract the input graph, shrinking it down until its size falls below a certain threshold.
Due to the project's size, we only focus on this component and compare an implementation based on dKaMinPar's application-specific MPI abstraction layer, a plain MPI-based implementation and a \kamping-based implementation.
We have extracted the shared code of all implementations (202 LOC) to a base class and only focus on the MPI-heavy part of the algorithm's implementation.
Here, the plain MPI-based implementation (154 LOC) is roughly 17.5\% larger than the \kamping-based implementation (127 LOC), which in turn is 16.5\% larger than the implementation based on dKaMinPar's specialized abstraction layer (106 LOC).
We observed the same running times for all variants.


\subsection{Integrating \kamping in RAxML-NG}
\label{sec:raxml-ng-integration}
As our largest application benchmark, we consider \mbox{RAxML-NG}.
RAxML~\cite{raxml} and its modern rewrite \mbox{RAxML-NG}~\cite{raxml-ng} are widely used real-world programs for phylogenetic inference in the field of bioinformatics with over \num{50000} citations.
\mbox{RAxML-NG} is written in \Cpp and uses a custom non-trivial abstraction layer over \texttt{pthreads} + MPI parallelism with over 700 lines of code.
We use \kamping to substantially simplify the MPI part of this abstraction layer; demonstrating that \kamping can easily be integrated even in large and well-established scientific codes which use hybrid parallelization (for an example, see \cref{fig:raxml-example}).
If \kamping had been available at the time, the RAxML-NG developers would have never needed to write, unit-test, maintain, and document over a hundred lines of complex code\footnote{Here, MPI and application code are heavily intertwined, making a fair comparison for exact LOC hard.}.

We empirically verified that replacing the abstraction layer with \kamping does not incur a measurable performance overhead even though RAxML-NG issued nearly 700 MPI calls per second\footnote{The mean running times are less than one standard deviations apart.}.
The binary's size does also not increase substantially (by \SI{2.5}{\percent}); the compilation time increases from 1:15\,min to 1:30\,min.

\begin{figure}[t]
	\begin{lstlisting}[
basicstyle=\ttfamily\scriptsize,
texcl,
% emph={}
emph={[2]send_recv_buf},
emphstyle={[2]\color{gray}},
emph={[5]as_serialized},
emphstyle={[5]\color{my-teal}},
escapebegin=\;\color{my-dark-red},
emph={[3]vector, std, exclusive_scan},
emphstyle={[3]\color{my-purple}},
tabsize=2
% TODO Highlight MPIFailureDetected?
]
// Before. The self-written mpi\_broadcast(...) wrapper and
// serialization/deserialization of data is not shown.
template<typename T>
static void mpi_broadcast(T& obj) {
  if (_num_ranks > 1) {
    size_t size = master() ?
      BinaryStream::serialize(
        _parallel_buf.data(),
        _parallel_buf.capacity(),
        obj)
      : 0;
    mpi_broadcast((void *) &size, sizeof(size_t));
    mpi_broadcast((void *) _parallel_buf.data(), size);
    if (!master()) {
      BinaryStream bs(_parallel_buf.data(), size);
      bs >> obj;
    }
  }
}

// After. \kamping provides all required functionality.
template <typename T>
static void mpi_broadcast(T &obj) {
  if (_num_ranks > 1) {
    _comm->bcast(send_recv_buf(as_serialized(obj)));
  }
}
\end{lstlisting}
  \caption{
    Example of a routine in the RAxML-NG parallelization abstraction layer simplified substantially using \kamping. We are able to replace custom serialization logic entirely.
  }
  \label{fig:raxml-example}
\end{figure}

\section{Towards General Building Blocks for Distributed Computing}
\label{sec:general-building-blocks}
Ease of development for MPI applications can be massively improved by providing a standard library of distributed (communication) algorithms and data structures, but incorporating this functionality in \kamping's core would make it overly complex.
With \kamping we ship multiple library extensions (plugins) including an STL-like distributed sorter (see \cref{sec:sorting}), specialized personalized \alltoall{} communication, fault-tolerance, and reproducible reduction operations which we will briefly highlight in the following.

\subsection{Sparse and Low-Latency All-To-All communication}
\label{sec:sparse-alltoall}
\Alltoall{} exchanges are one of the most frequent communication patterns in distributed computing.
However, there is a large algorithmic design space for \alltoall{} communication ranging from algorithms with near-optimal communication volume but latency at least linear in the number of processing elements to algorithms following a hypercube communication scheme with logarithmic latency but also a communication volume that is increased by a logarithmic factor \cite{Sanders2019}.

With \kamping's \texttt{GridCommunicator} plugin we go part of the way of trading communication volume for reduced latency by resorting to two-dimensional grid communication \cite{Kale2003}.
The processors are organized in a virtual two-dimensional grid and messages are routed in two hops to their destination to achieve a message start-up latency in $\mathcal{O}(\sqrt{p})$, where $p$ denotes the size of the communicator.
This enables hardware-agnostic latency reduction with asymptotic guarantees, in contrast to the variants provided by most MPI implementations.

Additionally, MPI's standard collectives have not been designed with \emph{sparse} communication patterns in mind.
\MPI{Alltoallv}, for example, requires a send counts parameter consisting of an array with one entry for each rank of the communicator, yielding a time complexity linear in the communicator size.
To mitigate this problem for static communication patterns, \emph{neighborhood} collectives have been added to MPI-3.0 allowing the user to perform \MPI{alltoall(v)} and \MPI{allgather(v)} on a previously defined (sparse) graph topology.
However, for applications and algorithms with rapidly changing communication partners, e.g., (dynamic) graph algorithms, the overhead of defining a new communication graph topology every few \alltoall{} exchanges may impose too much overhead.
\kamping's \texttt{SparseAlltoall} plugin offers a lightweight alternative for these scenarios, accepting a set of destination-message pairs as argument.
For the actual data exchange, the plugin uses the \emph{NBX} algorithm for sparse \alltoall{} communication by Hoefler~et~al.~\cite{Hoefler2010}.

Both techniques for improving irregular sparse exchanges work especially well for improving scalability of distributed graph algorithms~\cite{Sanders2023, Sanders2023b}.
\cref{plot:bfs-benchmark} shows an evaluation of our different \alltoall{} strategies using the weak-scaling BFS benchmark introduced earlier on three different graph families~\cite{Funke2019}, where each rank holds $2^{12}$ vertices and $2^{15}$ edges.
\ErdosRenyi{} graphs possess almost no locality (most edges cross rank boundaries) but small diameter, whereas random geometric graphs (RGG) are highly local with a high diameter.
Regarding locality, random hyperbolic graphs (RHG) range somewhere in between and also have small diameter. In contrast to \ErdosRenyi{} graphs and RGGs they possess high-degree vertices.
In the experiment, we compare different algorithms for the actual frontier exchange in each BFS step: built-in \MPI{Alltoallv} (\texttt{mpi, \kamping}), \kamping's sparse \alltoall{} plugin, \kamping's grid \alltoall plugin, and \MPI{Neighbor\_alltoallv} (\texttt{mpi\_neighbor}).
For RHGs (and less pronounced for \ErdosRenyi{} graphs) the most scalable communication method is our grid \alltoall{} approach.
Grid \alltoall also outperforms built-in \MPI{Alltoallv} on RGGs.
Due to their high diameter and locality, a competitive performance on RGGs can only be achieved with sparse communication.
\kamping's sparse \alltoall approach is only slightly slower than \MPI{Neighbor\_alltoallv}.
Note that when rebuilding MPI's communication graph before each frontier exchange, which simulates dynamic communication patterns to some extent, \MPI{Neighbor\_alltoallv} does not scale.

\subsection{User-Level Failure Mitigation}
\label{sec:fault-tolerance}

With the increasing number of processors in high-performance computing clusters, the probability that some processors fail during a computation rises.
Handling such failures constitutes a major challenge for future exascale systems~\cite{Shalf2010}.
In upcoming systems, we expect a hardware failure to occur every 30 to 60 minutes~\cite{Cappello2014,Dongarra2015,Snir2014}.
The upcoming MPI~5.0 standard enables developers to develop software able to recover from such failures by employing User-Level Failure-Mitigation (ULFM)~\cite{ulfm}.
As one example of the potency of our plugin architecture (\cref{sec:expandability}) we developed an abstraction layer over ULFM supporting all functions of the proposal\footnote{\url{https://fault-tolerance.org/}}.
As plugins can add custom error handling hooks, this enables users to develop fault-tolerant algorithms using idiomatic \Cpp exceptions instead of checking return codes,
as shown in \cref{fig:ulfm-example}.
\begin{figure}
	\begin{lstlisting}[
          basicstyle=\ttfamily\scriptsize,
          emph={shrink, revoke, allreduce, MPIFailureDetected, is_revoked},
emph={[2]send_buf, send_counts},
emphstyle={[2]\color{gray}},
emph={[3]vector, std, exclusive_scan},
emphstyle={[3]\color{my-purple}}
]
try {
  comm.allreduce(/* ... */);
} catch ([[maybe_unused]] MPIFailureDetected const& _) {
  if (!comm.is_revoked()) {
      comm.revoke();
  }
  // Create a new communicator containing only the
  // surviving processes.
  comm = comm.shrink();
}
\end{lstlisting}
	\caption{
		Handling a process failure using \kamping's ULFM plugin.
	}
	\label{fig:ulfm-example}
\end{figure}

\subsection{Reproducible Reduce}
\label{sec:reproducible-reduce}

\begin{figure}[t]
    \centering
    \includegraphics[width=0.35\textwidth]{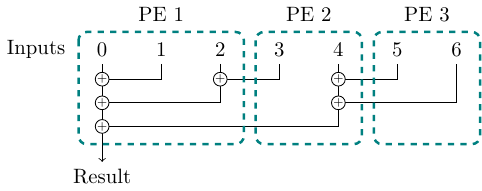}
    \caption{Reduction tree for 7 elements distributed across 3 ranks.
    The operations are performed in the order described by the tree.}
    \label{fig:reduction-tree}
\end{figure}

Reproducibility of results is a key aspect of scientific work.
One challenge in distributed computing is to ensure that the results of a computation are consistent across different runs using different numbers of processors. IEEE~754 floating point math is not associative because of rounding errors and thus the order of operations, which often depends on the number of processors, influences the result.

We include an implementation of \texttt{MPI\_Reduce} as a \kamping plugin which fixes the reduction order independent of the number of processors but is faster than a gather + local reduction + broadcast.
We use a binary tree scheme inspired by Villa~et~al.~\cite{Villa2009} and include various performance improvements (\cref{fig:reduction-tree}; for details see Stelz~\cite{Stelz2022}).
This enables parallel computations while using only a few messages to exchange intermediate results.

Similar to \enquote{normal} \kamping \texttt{reduce}, we support plain-MPI constants, function pointers, and lambda functions as operations.
We hope that the availability of a convenient off-the-shelf library method encourages more authors to ensure that their applications produce consistent and reproducible results.

\section{Conclusion}
\label{sec:conclusion}
We introduced \kamping, a set of novel near zero-overhead \Cpp MPI bindings.
Through configurable inference of parameter defaults, fine-grained memory allocation control, enhanced safety guarantees, and a flexible plugin system, it enables both rapid prototyping and careful engineering of distributed algorithms, which we demonstrated using a variety of benchmarks from different application domains.

\kamping is open source, extensively tested, and currently used in multiple research projects.
In the future, we plan to extend the standard coverage while also working towards our goal of building a basic algorithmic toolbox on top of it, to ease rapid prototyping and analysis of distributed algorithms with a strong focus on performance.
To this end, we will also further elicit the needs of the MPI community.

We are currently working on generalizing the indirection patterns for \alltoall{} primitives to higher dimensions, while also incorporating message aggregation. This is applicable in both request-replay patterns when reading from globally distributed data, and algorithms with highly-irregular communication without hard synchronization. We are implementing these building blocks on top of \kamping and will integrate them into future library releases.

With distributed containers, we want to enable lightweight bulk parallel computation inspired by MapReduce~\cite{Dean2004} and Thrill~\cite{Bingmann2016}, while not locking the programmer into the walled garden of a particular framework.
We strive to establish \kamping as a stable core for a whole ecosystem of future general-purpose distributed algorithms and applications.

\section*{Acknowledgment}
The authors gratefully acknowledge the Gauss Centre for Supercomputing e.V. (\url{www.gauss-centre.eu}) for funding this project by providing computing time on the GCS Supercomputer SuperMUC at Leibniz Supercomputing Centre (\url{www.lrz.de}).

This work was performed on the HoreKa supercomputer funded by the Ministry of Science, Research and the Arts Baden-Württemberg and by the Federal Ministry of Education and Research.

ChatGPT-3 was used to verify the grammatical correctness of a handful of sentences.
No generated text has been used in the manuscript.
GitHub Copilot was used to assist in writing unit tests and code documentation, and for minor code-completion.
No code generation has been used for interface, library, and algorithm design.

We thank Marvin Williams for helping us with his \Cpp expertise and assistance in \Cpp standard-related matters.


\IEEEtriggeratref{34}
\bibliographystyle{IEEEtran}
\bibliography{IEEEabrv, IEEE_etal,references.bib}

\end{document}